\documentclass[12pt, draftclsnofoot, onecolumn]{IEEEtran}

\usepackage{cite}
\usepackage{amsmath,amssymb,amsfonts}
\usepackage{algorithmic}
\usepackage{graphicx}
\usepackage{epstopdf}
\usepackage{textcomp}
\usepackage{xcolor}
\usepackage{subfigure}
\usepackage{amsthm}
\usepackage{mathrsfs}

\usepackage{pdfpages}
\usepackage{colortbl}
\definecolor{mygray}{gray}{0.85}
\usepackage{array}
\usepackage{caption}
\usepackage{setspace}
\usepackage{diagbox}
\definecolor{mygray}{rgb}{0.96,0.99,0.96}
\definecolor{mypink}{rgb}{.99,.93,.85}
\definecolor{mycyan}{cmyk}{.2,0.04,0,0}
\usepackage[linesnumbered,boxed,ruled,commentsnumbered]{algorithm2e}
\usepackage{xcolor}  
\usepackage{colortbl,booktabs}
\usepackage{multirow}
\usepackage[ruled,linesnumbered]{algorithm2e}

{}
{}
{}

\newcolumntype{I}{!{\vrule width 1.25pt}}
\newlength\savedwidth

\newlength\savewidth
\newcommand\shline{\noalign{\global\savewidth\arrayrulewidth
		\global\arrayrulewidth 1.25pt}%
	\hline
	\noalign{\global\arrayrulewidth\savewidth}}

\usepackage{amsmath,bm}
\usepackage{cite}
\usepackage{amsmath,amssymb,amsfonts}
\usepackage{algorithmic}
\usepackage{amsthm}
\usepackage{graphicx}
\usepackage{textcomp}
\usepackage{xcolor}

\begin{document}

\title{\Huge AI-Assisted MAC for Reconfigurable Intelligent Surface-Aided Wireless Networks: \\Challenges and Opportunities}

 \author{
\IEEEauthorblockN{Xuelin Cao, Bo~Yang,~\IEEEmembership{Member,~IEEE},
 Chongwen Huang,~\IEEEmembership{Member,~IEEE},\\ 
 Chau Yuen,~\IEEEmembership{Fellow,~IEEE},  
 Marco Di Renzo,~\IEEEmembership{Fellow,~IEEE},\\
 Zhu Han,~\IEEEmembership{Fellow,~IEEE},
 Dusit Niyato,~\IEEEmembership{Fellow,~IEEE},\\
 H. Vincent Poor,~\IEEEmembership{Life Fellow,~IEEE}, and
 Lajos Hanzo,~\IEEEmembership{Fellow,~IEEE}
 }
  \thanks{X. Cao, B. Yang, and C. Yuen are with the Engineering Product Development Pillar, Singapore University of Technology and Design, Singapore. (e-mail: xuelin$\_$cao, bo$\_$yang, yuenchau@sutd.edu.sg).}%
\thanks{C. Huang is with the Zhejiang Provincial Key Lab of information processing, communication and networking, Zhejiang University, P.R. China. (e-mail: chongwenhuang@zju.edu.cn).}
\thanks{M. Di Renzo is with Universit\'e Paris-Saclay, CNRS, CentraleSup\'elec, Laboratoire des Signaux et Syst\`emes, 3 Rue Joliot-Curie, 91192 Gif-sur-Yvette, France. (e-mail: marco.di-renzo@universite-paris-saclay.fr)}
\thanks{Z. Han is with the Department of Electrical and Computer Engineering, University of Houston, USA. (e-mail: zhan2@uh.edu).}
\thanks{D. Niyato is with the School of Computer Science and Engineering, Nanyang Technological University, Singapore. (e-mail: dniyato@ntu.edu.sg).}
\thanks{H. V. Poor is with the Department of Electrical Engineering, Princeton University, USA. (e-mail: poor@princeton.edu).}
\thanks{L. Hanzo is with the School of Electronics and Computer Science, University of Southampton, U.K. (e-mail: lh@ecs.soton.ac.uk).}
}
\maketitle

\begin{abstract}
Recently, significant research attention has been devoted to the study of reconfigurable intelligent surfaces (RISs), which are capable of reconfiguring the wireless propagation environment by exploiting the unique properties of metamaterials-based integrated large arrays of inexpensive antennas. Existing research demonstrates that RISs significantly improve the physical layer performance, including the wireless coverage, achievable data rate and energy efficiency. However, the medium access control (MAC) of multiple users accessing an RIS-enabled channel is still in its infancy, while many open issues remain to be addressed. In this article, we present four typical RIS-aided multi-user scenarios with special emphasis on the MAC schemes. We then propose and elaborate upon centralized, distributed and hybrid artificial-intelligence (AI)-assisted MAC architectures in RIS-aided multi-user communications systems. Finally, we discuss some challenges, perspectives and potential applications of RISs as they are related to MAC design.
\end{abstract}

\IEEEpeerreviewmaketitle

\section{Introduction}
With the ever-increasing demands on wireless networks, research in wireless communications continues to focus on meeting the challenges of improving the energy efficiency (EE) versus spectral efficiency (SE) trade-offs. Advances in meta-materials have fuelled research in reconfigurable intelligent surfaces (RISs) for beneficially reconfiguring the wireless communication environment with the aid of a large array of low-cost reconfigurable elements. This new design paradigm results in migration from traditional wireless connections to ``intelligent-and-reconfigurable connections". The intelligently controlled features of RISs lead to potential benefits for future wireless networks, such as their coverage enhancement, EE/SE performance improvement, leading to improved throughput and security \cite{di2019smart}. Because of these potential benefits, RISs are eminently suitable for addressing various challenges of wireless communications; hence they have been extensively investigated in diverse applications. Although the benefits of RISs in the physical layer have already been validated in practice, their performance is still constrained by the medium access control (MAC) layer, since the real-time configuration of RISs is complex and hence costly. To address this problem and improve the benefits of RISs, artificial intelligence (AI)-based methods can be applied to design MAC protocols for RIS-aided wireless networks.

\begin{figure*}[t]
\centering{\includegraphics[width=6.5in,height=3in]{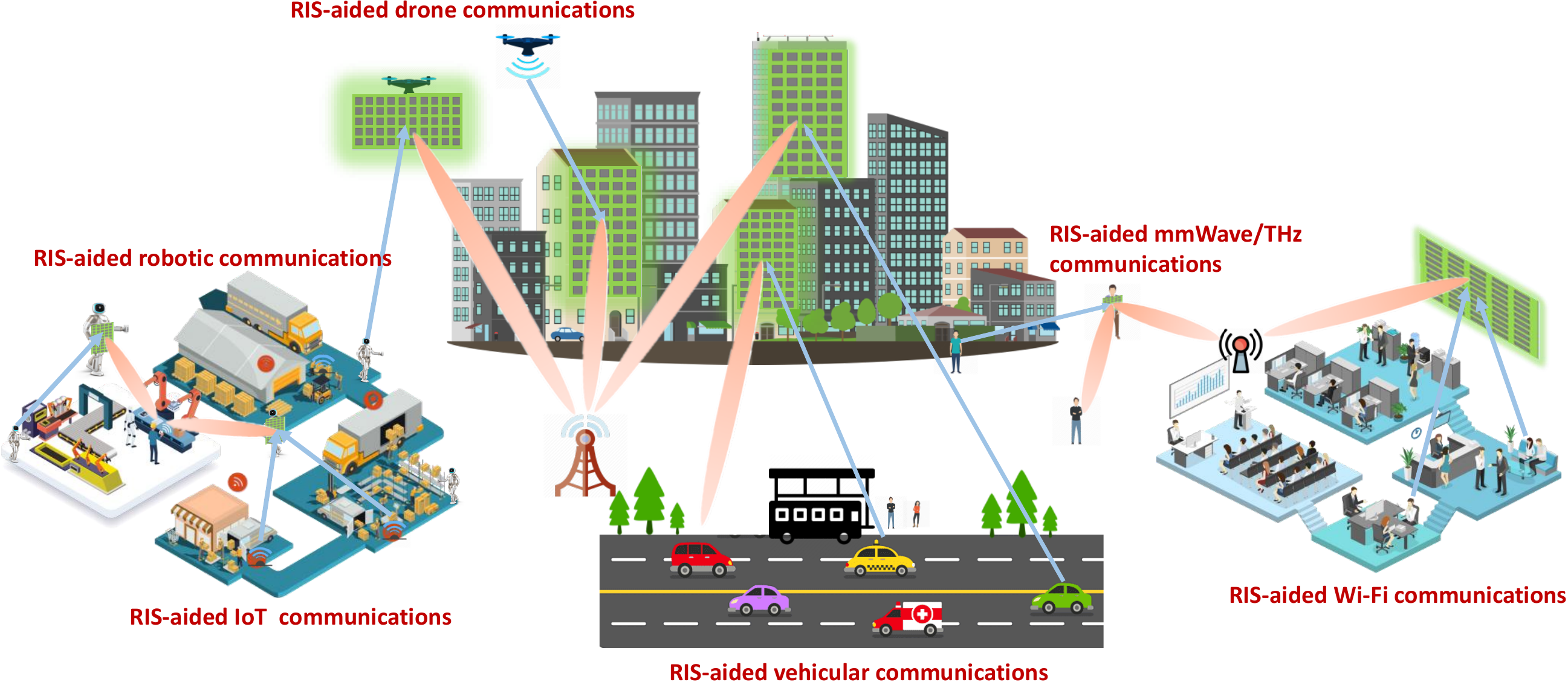}}
	\caption{\small RIS-aided wireless network applications in multi-user communication systems.}
	\label{App}
\end{figure*} 

Most of the existing research activities on RISs focus on physical layer issues, such as the issues of RIS deployment and their sustainable operation, flexible beamforming reconfiguration, EE/SE performance improvement and their compatibility with emerging technologies such as non-orthogonal multiple access (NOMA), as well as massive multiple-input multiple-output (MIMO) aided millimeter-wave (mmWave)/terahertz (THz) communications \cite{SGong}. Following the recent breakthrough in the fabrication of programmable metamaterials, RISs have been employed in various wireless networks. Some of the MAC-related issues have also been investigated in order to support seamless connectivity \cite{YYang, SLin, CYou, JHu, SLi, TBai, MJung, HGuo, ZDing, GYang, XCao}. However, these explorations of the MAC layer have focused primarily on the single-user uplink (UL) or multi-user downlink (DL). But there is a paucity of promising multi-user uplink solutions, since this scenario has not attracted significant attention to date. With the continued development of RIS technology and its integration with AI, the latency-sensitive services/applications supported by RISs are of salient importance in the 6G area, including RIS-aided vehicular, drone, or robotic communications, as shown in Fig. \ref{App}. Moreover, handling the massive number of RIS-aided sensors or Internet-of-things (IoT) devices represents a significant challenge in terms of the EE/SE. Finally, RISs are eminently suitable for supporting the emerging mmWave/THz communications in pursuit of high quality of service (QoS). Clearly, compelling MAC designs have to be conceived for fully exploiting the potential of RIS-aided wireless networks. 


Against this background, we first present four typical scenarios (\textbf{S}) of RIS-aided multi-user wireless communications, with special emphasis on their MAC protocol design. Then, we propose three types of AI-assisted MAC structures designed for the RIS-aided multi-user uplink and discuss their protocols and applications. Next, we discuss some potential challenges facing AI-assisted MAC protocols. Furthermore, we evaluate the proposed AI-assisted MAC solutions to quantify their system throughput. Finally, we conclude with the trends in designing AI-assisted MAC protocols for RIS-aided wireless networks.

\section{Scenarios, Protocols and Objectives}
\begin{figure*}[t]
\centering{\includegraphics[width=4.2in,height=2.6in]{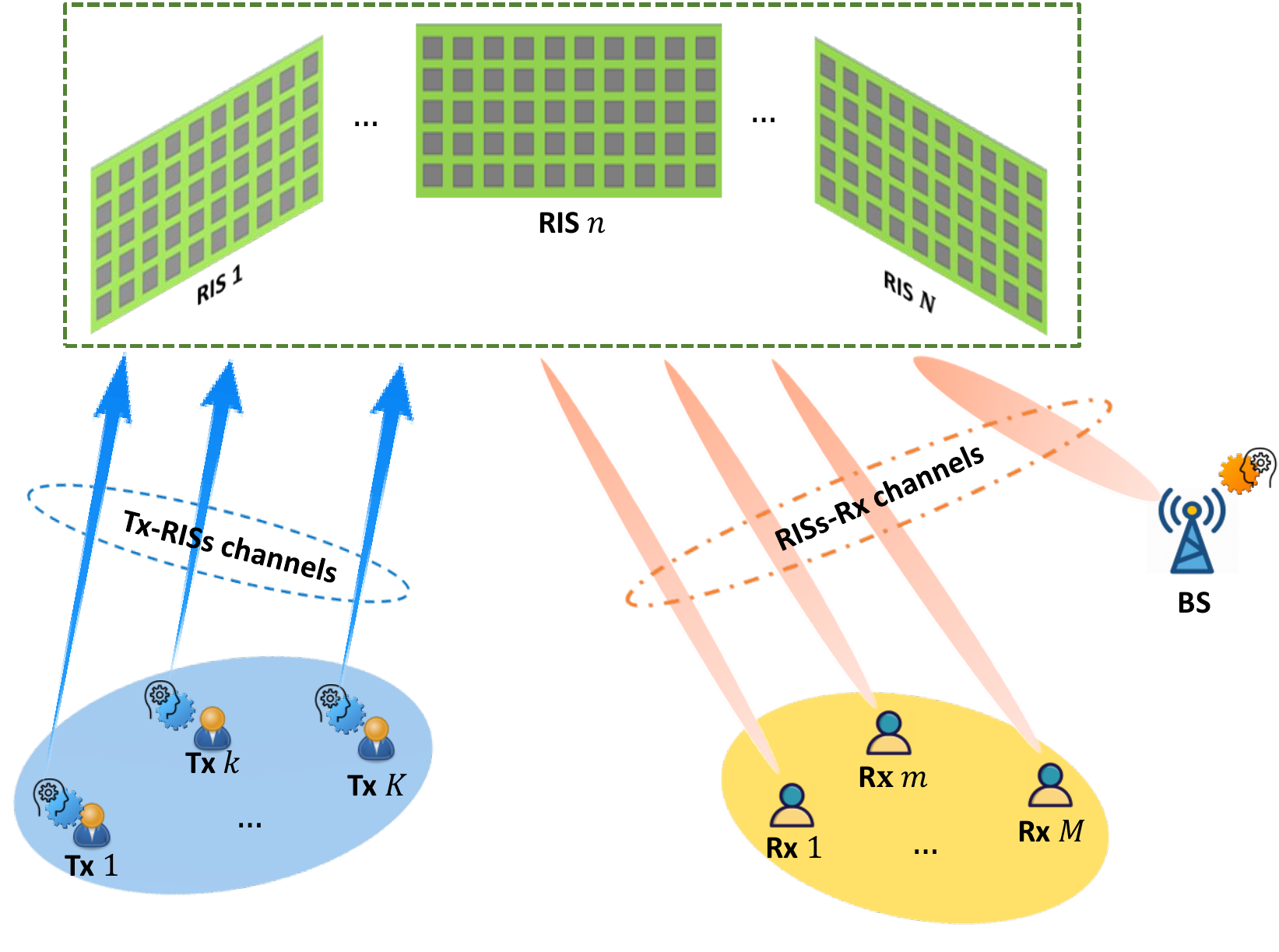}}
	\caption{\small AI-assisted MAC scenarios in RIS-aided wireless communications.}
	\label{Sec}
\end{figure*}

In this section, we commence by presenting our typical MAC scenarios seen in Fig. \ref{Sec} and then review the existing MAC protocols and their design objectives.  


\subsection{Scenarios}
\begin{itemize}
\item[-] \textbf{S1: Single RIS-aided multiple-Tx single-Rx}. In S1, a single RIS is deployed to coordinate the uplink transmissions of multiple transmitters (Txs) to a base station (BS), e.g., $K$ Txs to a BS via an RIS. Note that the single-input single-output scenario is a special case when $K=1$.

\item[-] \textbf{S2: Single RIS-aided multiple-Tx multiple-Rx}. In S2, a single RIS is deployed to coordinate the uplink transmissions of multiple Txs to multiple receivers (Rxs), e.g., $K$ Txs to $M$ Rxs via an RIS.  

\item[-] \textbf{S3: Multiple RIS-aided multiple-Tx single-Rx}. In S3, multiple RISs are coordinated to support the uplink transmissions of multiple Txs to a BS, e.g., $K$ Txs to a BS via $N$ RISs. 
 
\item[-] \textbf{S4: Multiple RIS-aided multiple-Tx multiple-Rx}. In S4, multiple RISs are coordinated to assist the uplink transmissions of multiple Txs to multiple Rxs, e.g., $K$ Txs to $M$ Rxs via $N$ RISs.  
\end{itemize}

For future networks with the ultra dense deployment of users, the coordination of massive users for meeting their QoS demands in the scenarios with a single RIS (e.g., S1 and S2) becomes challenging. For the scenarios having multiple RISs (e.g., S3 and S4), the user-RIS association (also known as RIS allocation) becomes more attractive, since serious interference  may occur among the RISs.

\subsection{MAC Protocols}

Conceiving MAC protocols (\textbf{P}) for RIS-aided multi-user wireless communications have become essential. Both the conventional orthogonal multiple access (OMA) and the emerging NOMA schemes have already been investigated \cite{YYang, SLin, CYou, JHu, SLi, TBai, MJung, HGuo, ZDing, GYang, XCao}. 
\par \textbf{RIS-aided OMA}. It aims for improving the SE/EE, for enhancing the QoS and for increasing the number of network connections by reconfiguring the wireless propagation environment\cite{YYang, SLin, CYou, JHu, SLi, TBai, MJung, HGuo}. The available OMA technologies integrated with RISs are enumerated as follows.
\begin{itemize}
\item[-] \textbf{P1: RIS-aided time division multiple access (TDMA)}. It enables multiple users to transmit their data via RISs on the same frequency in different time slots \cite{CYou, JHu, SLi, TBai, MJung, HGuo}. 
\item[-] \textbf{P2: RIS-aided frequency division multiple access (FDMA)}. It enables multiple users to transmit their information via RISs in the same time slot on non-overlapping domain frequency channels \cite{YYang, SLin, MJung}. 
\item[-] \textbf{P3: RIS-aided spatial division multiple access (SDMA)}. It enables multiple users to transmit their data via RISs either in unique angular direction or by exploiting the users' unique channel impulse responses (CIRs) by spatial multiplexing \cite{ZDing}. 
\item[-] \textbf{P4: RIS-aided carrier sensing multiple access (CSMA)}. It enables multiple users to transmit their signals via RISs relying on random contention-based multiple access protocols, where some control information exchange is required before the RIS-aided payload transmission \cite{SGong, XCao}. 
\end{itemize}
\par \textbf{RIS-aided NOMA}. With the assistance of RISs, NOMA schemes are capable of avoiding having to distinguish multiple users on the same resource block by their power levels, which may improve their SE/EE and latency. Hence, NOMA-assisted RIS-aided multi-user downlink communications have been explored in \cite{ZDing, GYang}, concluding that the performance of NOMA is not always preferable compared to OMA. For example, NOMA may perform worse than angularly-orthogonal SDMA or TDMA.

\subsection{Objectives}

The objectives (\textbf{O}) of MAC designs conceived for RIS-aided multi-user systems include the following potential aspects.  
\begin{itemize}
\item[-] \textbf{O1: System throughput}. The throughput is directly linked to the SE, which can be increased by increasing the number of transceivers and/or the time/frequency/space/RISs resources. 
\item[-] \textbf{O2: EE performance}. The EE is given by the capacity normalized by the MAC's energy consumption, which can be readily improved by the directional communications of RISs, which is mitigating the interference by directional beams and/or enhancing the strength of the desired signal reflected by passive elements. From a specific MAC design perspective, the EE can be further improved by avoiding access collisions. Therefore, a critical aspect of EE in the MAC is that of exploiting the angular focusing capabilities of RISs for multiple users.
\item[-] \textbf{O3: Fairness}. The rate fairness of different users should be guaranteed without degrading the overall system performance, especially when the resources are limited. In this context, avoiding these starvations of users suffering from low link quality becomes a pivotal criterion when designing an appropriate MAC for RIS-aided wireless networks. 
\item[-] \textbf{O4: Overhead}. The overhead of user access-grant directly affects both the communication and computational complexity in terms of the RIS channel estimation, reconfiguration and resource allocation. Additionally, the wireless handshake of the MAC design may impose extra costs. How to implement the MAC protocol at a low-cost in RIS-aided wireless networks is a challenging dilemma. 
\item[-] \textbf{O5: Latency}. The MAC design needs to meet the low-latency requirement of delay-sensitive applications with the aid of RISs.
\end{itemize}

Existing contributions on the MAC design of RIS-aided communication systems are summarized at a glance in Table \ref{SP},  with an emphasis on their design objectives and critical features. In the table, the acronyms SDR, STM, SCA, and BCD represent semidefinite relaxation, strongest tap maximization, successive convex approximation and block coordinate descent, respectively. Notably, with the context of the coming wireless intelligence era, AI will give impetus to the MAC designs in RIS-aided networks by enabling the users and BS to ``think-and-decide".
  
\begin{table*}[t] 
\newcommand{\tabincell}[2]{\begin{tabular}{@{}#1@{}}#2\end{tabular}}
		\small
		\centering
			\renewcommand{\arraystretch}{1.2}
			\captionsetup{font={small}} 
			\caption{\scshape RIS-Aided Transmission Protocol Design: State of the Art} 
			\label{SP}
			\footnotesize
			\centering  
			\begin{tabular}{|m{0.05\textwidth}<{\centering}|m{0.06\textwidth}<{\centering}|m{0.08\textwidth}<{\centering}| m{0.06\textwidth}<{\centering}|m{0.13\textwidth}<{\centering}|m{0.10\textwidth}<{\centering}|m{0.35\textwidth}| }  
				\shline
			    \rowcolor{mycyan} \textbf{Refs.} & \textbf{Scenarios} & \textbf{Protocols} &\textbf{Objectives} & \textbf{Approaches} & \textbf{Applications} & \textbf{Key features}\\
				\shline  
				\multicolumn{7}{|c|}{Centralized} \\ \hline
				\rowcolor{mypink}\cite{YYang} & \textbf{S1}  & \textbf{P1}, \textbf{P2} & \textbf{O1}, 
				\textbf{O2} & Iterative algorithm & IoT & An RIS-aided transmission protocol with the RIS group partition for long distance transmission.\\ 
			    \hline
			    \rowcolor{mypink}\cite{SLin} &  \textbf{S1}  & \textbf{P1}, \textbf{P2} &\textbf{O1} & SDR, STM & mmWave,  IoT & An RIS-aided transmission protocol with frequency-selective channels for delay-sensitive applications.\\ 
			   \hline
			    \rowcolor{mypink}\cite{CYou} & \textbf{S1}  & \textbf{P1} & \textbf{O1} & Iterative algorithm & Wi-Fi & A pilot-assisted block transmission protocol for RIS channel estimation and passive beamforming with discrete phase-shift.\\ 
			    \hline
			    \rowcolor{mypink}\cite{JHu} & \textbf{S1}, \textbf{S3}  & \textbf{P1} & \textbf{O4} & Iterative algorithm, supervised learning  & RF sensing &  A periodic configuring protocol aims to perform RIS-aided human posture.\\ 
			    \hline
			    \rowcolor{mypink}\cite{TBai} & \textbf{S1}  & \textbf{P1} & \textbf{O5} & Iterative algorithm (BCD) & Edge computing & RIS aided TDD transmission is investigated in mobile edge computing systems.\\ 
			    \hline
			   \rowcolor{mypink} \cite{SLi} & \textbf{S1}  & \textbf{P1} & \textbf{O1} & Iterative algorithm (SCA) & Drone communications & Jointly optimizing UAV trajectory and RIS beamforming for RIS-aided UAV communications.\\ 
			    \hline
			   \rowcolor{mypink} \cite{MJung} & \textbf{S3} & \textbf{P1}, \textbf{P2} & \textbf{O1}, \textbf{O2} & Iterative algorithm &  IoT, massive & RIS-aided transmission combined TDD with OFDMA is discussed by jointly considering the user scheduling and power control.\\
			  \hline
			  \rowcolor{mypink}  \cite{HGuo} &  \textbf{S1}  & \textbf{P1} & \textbf{O1}, \textbf{O2} & Alternating optimization (BCD) & Wi-Fi & RIS-aided TDD transmission with considering perfect channel state information (CSI) and imperfect CSI. \\ 
			   \hline
			  \rowcolor{mypink}  \cite{{ZDing}} & \textbf{S3}  & \textbf{P3}, \textbf{NOMA} & \textbf{O1} &  Cauchy-Schwarz inequality & Edge computing & RIS-aided NOMA transmissions are used to serve multiple users on each orthogonal spatial direction. \\ 
			   \hline
			 \rowcolor{mypink}   \cite{GYang} & \textbf{S1}  & \textbf{NOMA} & \textbf{O1}, \textbf{O2} & Iterative algorithm (BCD, SDR) & Massive, low-latency & An RIS-aided NOMA with combined-channel-strength is proposed while ensuring the fairness among users.\\ 
			  \hline
			  \multicolumn{7}{|c|}{Distributed} \\ \hline
			  \rowcolor{mygray}\cite{di2019smart,SGong} & \textbf{S1}-\textbf{S4}  &  $\bf{\times}$ & \textbf{O1}-\textbf{O3} & $\bf{\times}$ & Wi-Fi, D2D, mmWave, IoT & Randomly access with information exchange in RIS-aided multi-user system.\\ \hline
			   \rowcolor{mygray}\cite{XCao} & \textbf{S1}, \textbf{S2}  & \textbf{P1}, \textbf{P2}, \textbf{P4} & \textbf{O1} & Alternating optimization & IoT, D2D, Wi-Fi & Randomly reserved access in RIS-aided multi-user system.\\ 
			    \shline

			\end{tabular}  
	\end{table*}

\begin{figure*}[t]
\centering{\includegraphics[width=6in,height=2.9in]{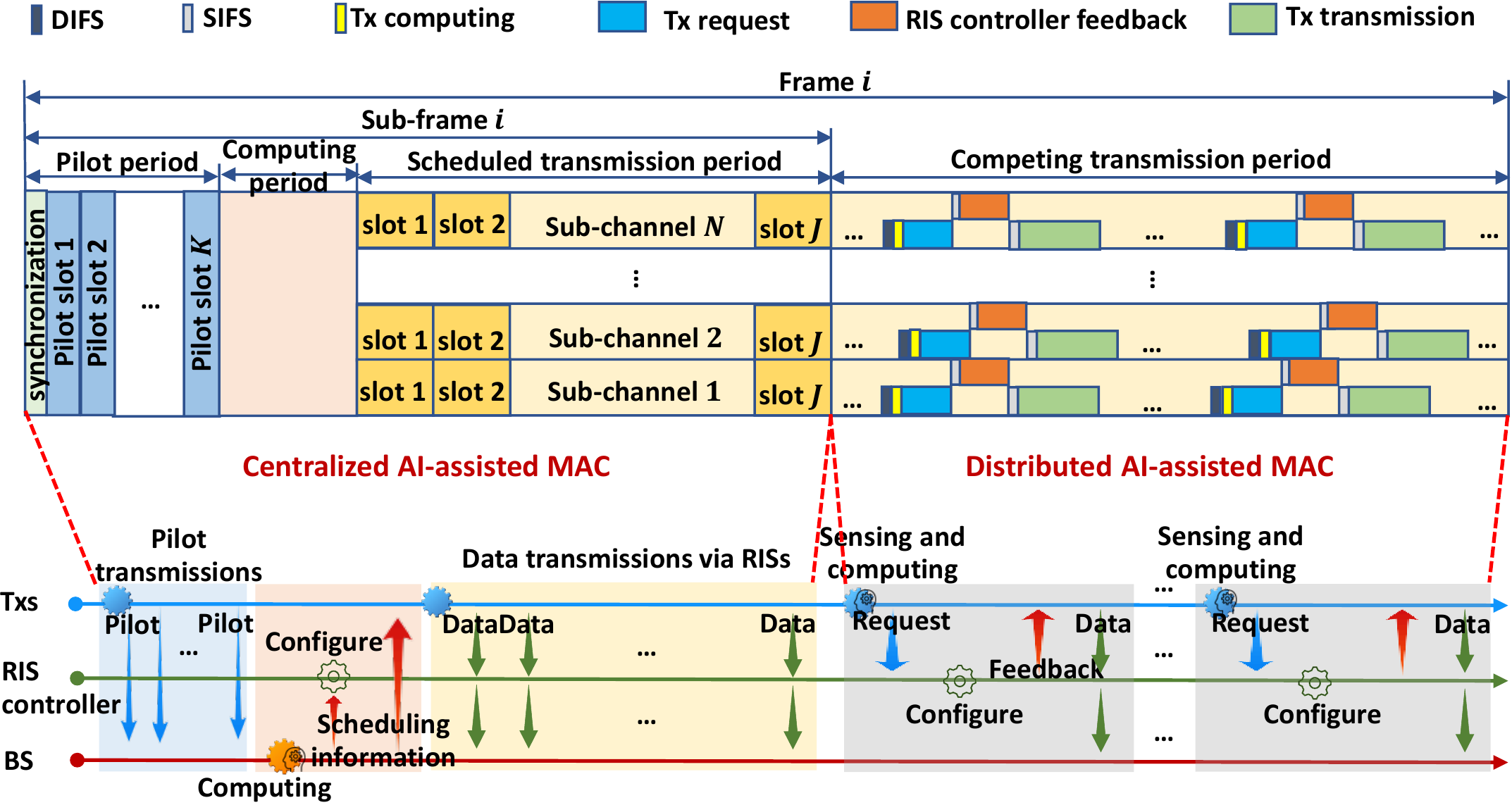}}
	\caption{\small Hybrid AI-assisted MAC structure
	 with centralized and distributed designs, an illustration of Case 1.}
	\label{PL}
\end{figure*}

\section{AI-Assisted MAC for RIS-Aided Networks}
In this section, we design three AI-assisted MAC architectures for RIS-aided wireless networks, namely for centralized, distributed and hybrid schemes. We discuss their differences from the perspective of their overall framework, protocol design, computational aspects and promising applications. In particular, the centralized and hybrid MAC protocols are time-frame based, while the distributed MAC protocol is based on a random access scheme. The centralized MAC design obeys a central schedule, the distributed MAC design uses a contention-based model, and the hybrid MAC design combines both features. Additionally, deep learning and reinforcement learning solutions are adapted to the different MAC protocols.  

\subsection{Centralized AI-Assisted MAC}
\subsubsection{Framework} In the proposed centralized AI-assisted MAC framework, the BS tightly coordinates the multiple access of users. Explicitly, the BS enables each of the RIS-controllers to beneficially configure the wireless propagation environment via deep learning for multiple users. Here, the RISs are assumed to be passive, simply reflecting the incident signals without sensing or processing. More explicitly, the BS has to estimate the concatenated BS-RIS-user link, calculate the RIS phase reconfiguration, and allocate resources via deep learning, as it will be detailed below in the `computation' part. As shown in Fig. \ref{PL}, each sub-frame is divided into three periods: the pilot period, computing period and scheduled transmission period. The pilot period and scheduled transmission period can be further divided into $K$ pilot slots and $J$ data slots, respectively. The users transmit their data in the $J$ data slots over the $N$ non-overlapping sub-channels. Based on this, the centralized AI-assisted MAC protocol is designed by giving full cognizance to both channel estimation as well as phase computation and data transmission.  
\subsubsection{Protocol} The protocol of the centralized AI-assisted MAC is shown in Fig. \ref{PL}, which is integrated with TDMA and FDMA, where each user obeys the time division scheme in each sub-channel. In particular, after synchronization, each user initiates pilot transmission to the BS in dedicated pilot slots. During the computational period, the BS first estimates the concatenated RIS link, followed by time, frequency and power resources allocation and RIS phase configuration. Then the BS instructs each RIS-controller to configure its reflection parameters and schedules the access of users, who transmit their data to the BS via the RISs.
\subsubsection{Computation}
Given the excessive complexity of high-dimensional full-search-based centralized MAC protocols, a deep learning-based computational model trained offline can be employed at the BS for finding a near-optimal solution at a reduced complexity. The input of the trained deep learning model can be the number of users, the number of sub-frames, the number of channels, the number of RISs and the RIS channel information. The online inference that moves the complexity to offline training is performed at the BS, which relies on a model-based training for determining the RIS phase-shift configuration, the RIS-deployment strategy and the resource allocation strategy, which are related to each other. More explicitly, these related learning tasks share the same input parameters, thus learning multiple related tasks jointly improves the prediction accuracy and generalization capability compared to learning them separately.

\subsubsection{Applications}
Given the centralized implementation and deep learning-based computation model considered, the centralized AI-assisted MAC design advocated can be readily applied to the scenarios S1 and S3 for supporting low-power RIS-aided communications.

\subsection{Distributed AI-Assisted MAC}
\subsubsection{Framework} In contrast to the centralized scheme, in the proposed distributed AI-assisted MAC framework, each user configures the multiple access and computes the RIS configuration by itself based on the RIS-aided network environment. In this case, no BS assistance is necessary. Each RIS is assumed to be passive and to occupy a non-overlapping frequency channel. In contrast to the centralized MAC design, the user has to negotiate with the RIS-controller for channel access. The corresponding RIS-aided data transmission as illustrated in Fig. \ref{PL}. In particular, channel sensing and computation are carried out at the user side via reinforcement learning (RL) to determine the RIS configuration. Once a channel is idle, the user sends the RIS configuration information to the RIS-controller to negotiate the ensuing RIS-aided data transmission. 
 
\subsubsection{Protocol} The protocol of the proposed distributed AI-assisted MAC is shown in Fig. \ref{PL}, which is integrated with CSMA and FDMA, where each user follows the distributed coordination function (DCF) based scheme in each channel. In particular, a competing user senses the state of each sub-channel. Once a channel is sensed to be idle, the user contends for the access to the channel. Waiting for a DCF inter-frame space (DISF) and backoff, the user computes its RIS configuration based on RL and sends an RIS configuration request to the RIS-controller. If the RIS is available for the user, the RIS-controller configures its reflection parameters and sends its feedback to the user after a short inter-frame spacing (SIFS). Following the elapse of a SIFS, the user then transmits the data to the BS via the RIS. Note that the feedback from the RIS-controller is sent without a transmit radio frequency chain. Moreover, the access collisions of users can be alleviated by the RIS-controller. 
 
\subsubsection{Computation}
In the distributed MAC protocol, an RL-based computational model can be employed by each user to solve the resource allocation and RIS configuration problems because no RIS channel-information exchange is required. The RL model includes the following aspects: the current RIS configuration, the current RIS deployment, and the currently occupied resources (e.g., power) of each user. The model actions include three aspects: the update of RIS configuration, the motion-trajectory of the user and the occupied resources updates. The reward function is decided by the throughput requirement of the user. When the action taken by the user improves its data rate, the user obtains a positive reward. By contrast, for throughput reductions, the user receives a negative reward (also termed the penalty). The RL-based computational model is more suitable for small RISs to avoid potential dimensionality problems. If the RIS is large, the RIS elements may be partitioned into groups, where each group maintains the same RIS configuration.

\subsubsection{Applications}
Due to the distributed implementation and the RL-based computational model, the distributed AI-assisted MAC design can be applied to all the scenarios S1 to S4 for meeting low-latency requirements.

\subsection{Hybrid AI-Assisted MAC}
Based on the centralized and distributed MAC frameworks proposed, we now discuss three types of hybrid AI-assisted MAC designs, where the centralized and distributed implementations are integrated into a single frame. 
\subsubsection{Case 1} In this hybrid framework, the scheduled and the competing transmissions relying on RISs are combined after the pilot transmissions and computing, while enabling users to switch between them for meeting different QoS demands, as illustrated in Fig. \ref{PL}. Each frame is partitioned into four parts, namely, the pilot period, the computing period, the scheduled transmission period and the competing transmission period. According to the proposed distributed AI-assisted MAC design, the scheduled users transmit their data to the BS via RISs in the scheduled transmission period. After that, based on the distributed AI-assisted MAC design, the unscheduled users (i.e., the unserviced users that have sent their pilots and the new requesting users) transmit their data to the BS or the Rxs via RISs during the competing transmission period. Given the dynamic switching between two transmission modes, this scheme is capable of maintaining the target-rate, and it may be suitable for all scenarios (i.e., S1 to S4).
\subsubsection{Case 2} In this hybrid framework, the competing requests and the scheduled RIS-aided transmissions are combined into a single frame. Each frame consists of the competing request period, the computing period and the scheduled transmission period. The user sends a request to the BS and when a sub-channel becomes available during the competing request period, then the BS issues a feedback for acknowledgment. Based on the requests received, the BS controls the RISs and sends the scheduling information to users during the computing period. Afterwards, the scheduled users transmit their data to the BS via RISs in the scheduled transmission period. Due to the competitive access of Case 2, it can be applied in scenarios S1 and S3 for supporting RIS-aided smart homes or smart factories. 
\subsubsection{Case 3} This hybrid framework is similar to Case 2, since it combines the competing requests and the reserved RIS-aided transmissions into a single frame. The slight difference is that computing in Case 3 is carried out at the user, rather than at a BS or RIS-controller. When a sub-channel becomes idle, the user occupies the channel, then computes the required resources and RIS configuration based on RL and sends a request to the RIS-controller for reserving the resources for future RIS-aided transmissions. The RIS-controller sends a feedback to the user once a request is registered and controls the RIS. When the reserved transmission period arrives, the user transmits the data to the Rx or the BS via the RIS in the reserved slots. Since RL is mainly used for complexity reduction, Case 3 can be applied in all the scenarios S1 to S4 for supporting RIS-aided periodic communications. 

In practice, due to the implementation constraints, the phase shifts applied by the RIS elements belong to a discrete set. Hence, the RIS configuration may be viewed as a classification problem, which can be tackled by using deep learning. If the phase shifts can be configured continuously, e.g., for varactor-based RIS designs, the optimal configuration of the phase shifts of the RIS can be regarded as a regression problem.

\section{Challenges}
In this section, we discuss some potential research challenges of designing AI-assisted MAC protocols for RIS-aided wireless networks in terms of the availability of limited datasets, privacy and security, as well as spectrum sensing.
\subsection{Limited Datasets} 
Using conventional supervised learning for both resource allocation and RIS configuration requires complete labeled datasets. However, collecting sufficient training data is a challenge in practice. For this reason, semi-supervised learning, transfer learning, or even autonomous learning can be explored to overcome the challenge of limited datasets.
\subsection{Privacy and Security} 
The centralized learning employed for the MAC design of RIS-aided communications may rely on privacy-sensitive datasets. To address this potential vulnerability, distributed learning, e.g., federated learning \cite{WNi, KYang}, requires uploading of only local model parameters, thereby promoting data privacy, although steps must be taken to preserve privacy even with distributed learning. Moreover, privacy can be enhanced in both training and inference through the use of federated learning and neural network segmentation, respectively. Thus, distributed learning constitutes a promising solution for enhancing the privacy and security of MAC protocols in RIS-aided networks.
\subsection{Spectrum Sensing}
In practice, when designing an efficient AI-assisted MAC protocol for RIS-aided wireless networks, especially a distributed version, accurate spectrum sensing is quite critical for reducing collisions and interferences. Due to the typical hidden and exposed terminal problems, conventional carrier sensing techniques suffer from spectrum sensing deficiencies. To circumvent this impediment, intelligent spectrum learning relying on the cooperation of the RISs, BSs or Txs can be explored with the objective of improving the sensing efficiency and accuracy.

\begin{figure}[t]	
\centering{\includegraphics[width=0.44\textwidth]{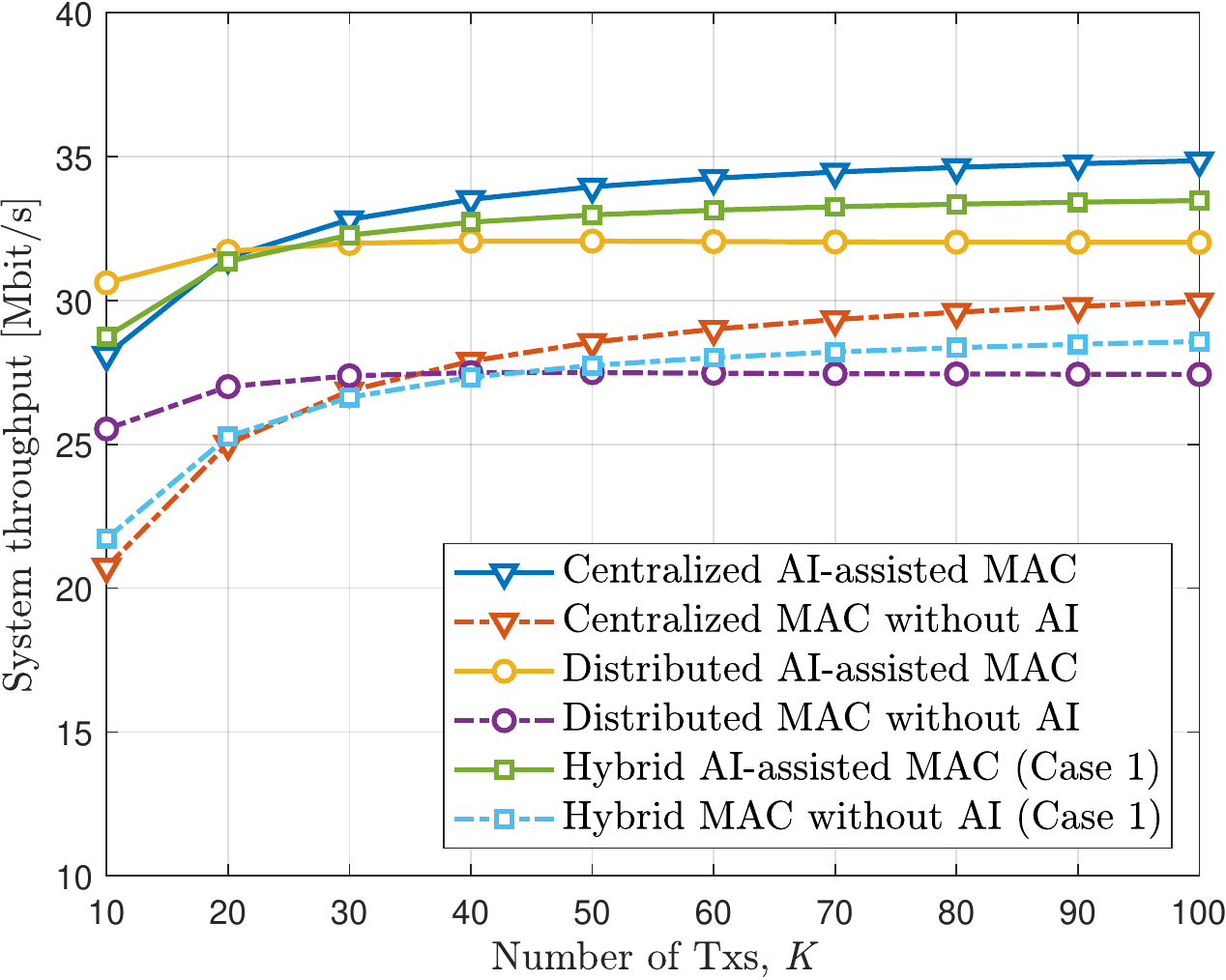}}
\caption{\small System throughput vs. the number of transmitters, where N=2.} 
\label{sim1}
\end{figure}

\section{Performance Evaluation}
This section evaluates both the system throughput and the EE of the three AI-assisted MAC frameworks proposed for RIS-aided wireless networks. We opt for Case 1 as the hybrid MAC protocol. We consider a scenario that consists of a BS, 2 RISs having 128 RIS-elements each, and 100 single-antenna Txs, where the Tx-RIS and RIS-BS distances are 50 m and 30 m, respectively. A Rician fading channel model is assumed, where the Tx-RIS and RIS-BS channels benefit from the existence of LoS links having a path loss exponent of 2.2, while the Tx-BS channels are NLoS links with a path loss exponent of 3.6. The power dissipated at each user is 10 dBm, the noise power is -94 dBm, the total bandwidth is 10 MHz and the number of sub-channels is 2. Furthermore, we assume that each RIS occupies a single sub-channel as a benefit of interference cancellation, and each Tx is only allowed to use a single RIS to communicate with the Rx at a time.

\subsection{System throughput vs. the number of Txs}
Figure \ref{sim1} shows the system throughput of RIS-aided wireless communications versus the number of Txs in the three types of AI-assisted MAC proposed. Firstly, it is observed that the system throughput of each AI-assisted MAC is improved compared to the MAC without AI, since AI methods have the potential of reducing the computing time. Observe from Fig. \ref{sim1} that the throughput of each MAC initially increases, but then tends to saturate as the number of Txs increases. This is because the computation time ratio within each frame is reduced upon increasing the length of each frame. Also, the system throughput of the distributed MAC exhibits a slight lesion after saturation due to the competition collisions. Additionally, the system throughput of the distributed MAC is best when the number of Txs is low (e.g., less than 20). As the number of Txs increases, the system throughput of centralized MAC becomes higher than that of the distributed MAC and the system throughput of hybrid MAC (Case 1) is in the middle. 
\subsection{EE performance vs. the number of Txs}
\begin{figure}[t]
\centering{\includegraphics[width=0.5\textwidth]{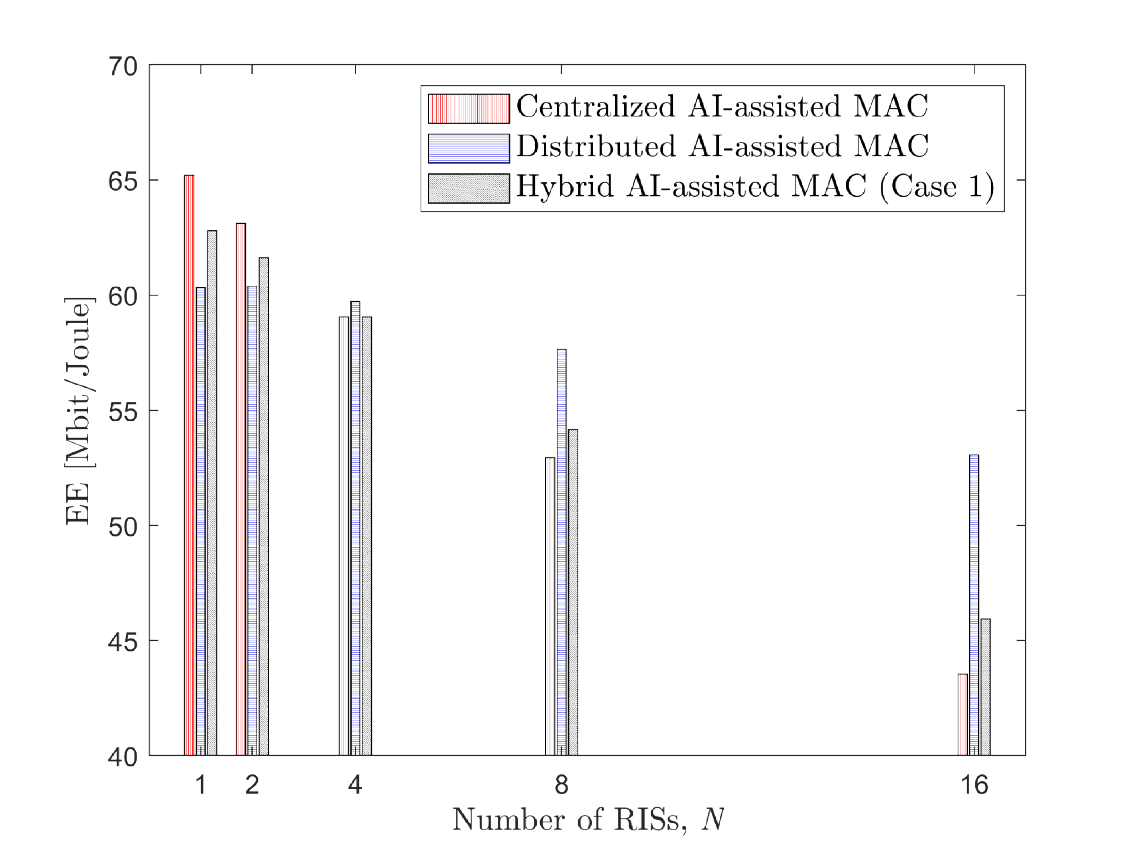}}
\caption{\small Energy efficiency vs. the number of RISs, where K=100.} 
\label{sim2}
\end{figure}

Figure \ref{sim2} shows the EE versus the number of RISs for all three types of AI-assisted MAC. It is observed that the EE of each type decreases as the number of RISs increases due to the increased computational complexity associated with extra computing time. Additionally, the EE of the centralized AI-assisted MAC is better than that of the distributed AI-assisted MAC, when the number of RISs is $1$ or $2$. As the number of RISs increases to $4$, $8$ or $16$, the EE of the distributed AI-assisted MAC has the edge, because the overhead imposed by the centralized AI-assisted MAC is increased. In other words, the centralized AI-assisted MAC is suitable for a small number of RISs, while the distributed AI-assisted MAC is recommended for a large number of RISs.

\section{Conclusion and Open Issues}\label{sec7}
In conclusion, we have presented four typical scenarios of RIS-aided multi-user communications. We then have reviewed the family of MAC solutions conceived for RIS-aided wireless networks and highlighted a range of competing MAC designs conceived for AI-assisted MAC structures relying on centralized, distributed and hybrid implementations. In particular, the centralized AI-assisted MAC excels in satisfying the target QoS of users, while the distributed AI-assisted MAC is more capable of meeting the random or unpredictable requirements of users. Finally, the family of hybrid AI-assisted MAC solutions strikes a beneficial trade-off between them. As performance evaluations revealed, distributed AI-assisted MAC schemes are more applicable to networks with small numbers of users associated with a large number of RISs. By contrast, the centralized AI-assisted MAC schemes are more suitable for a large number of users in conjunction with small numbers of RISs. 

We foresee open issues along the road of integrating AI-assisted MAC designs with next-generation technologies. For example, a community-effort is required for conceiving advanced AI computing models in MAC design, AI-assisted MAC solutions for both mmWave and THz communications, for privacy-preserving, for smart sensing, for controlling and optimizing AI-assisted MAC implementations. Promisingly, MAC designs with AI-empowered-RISs become more exciting and challenging. All in all, an exciting area for new research.

\end{document}